

\input amstex

\documentstyle{amsppt}
\font\deffont=cmbxti10
\define\mapright#1{\smash{\mathop{\to}\limits^{#1}}}

\define\mapdown#1{\Big\downarrow 
\rlap{$\vcenter{\hbox{$\scriptstyle#1$}}$}}

\define\mapleft#1{\smash{\mathop{\leftarrow}\limits^{#1}}}
\def\sd{\mathbin{ \raise0.0pt\hbox{ \vrule height5pt width.4pt 
depth0pt}\!\times}}
\define\reals{\text{{\rm I \! \!\! R}}}
\define\bDel{{{\Delta}}}
\define\bx{\text{{\bf x}}}
\define\bo{\text{{\bf 0}}}
\define\bu{\text{{\bf u}}}
\define\tilM{\widetilde M} 
\define\q{{\eufm q}}
\define\qk{{\eufm q_{\sssize \kappa}}}
\define\qM{{\eufm q_{\sssize M}}}
\define\qE{{\eufm  q_{\sssize  E}}}
\define\qi{{i}}
\def\qed{\hfill\hbox{\vrule width 4pt height 6pt depth 1.5 pt}}

\define\boxit#1{\vbox{\hrule\hbox{\vrule\kern3pt\vbox{\kern3pt#1\kern3pt}
\kern3pt\vrule}\hrule}}

\NoBlackBoxes

\topmatter
\title Group Invariant Solutions in Mathematical Physics and 
Differential Geometry \endtitle

\rightheadtext{Group Invariant Solutions \hfil}

\author I.M. Anderson, M. E. Fels, C.G. Torre \endauthor
\address Dept. of Mathematics and Statistics, Utah State University, 
Logan Utah, 84322 \endaddress  
\email anderson\@math.ams.org\endemail
\address Dept. of Mathematics and Statistics, Utah State University, 
Logan Utah, 84322 \endaddress  
\email fels\@math.ams.org\endemail
\address Dept. of Physics, Utah State University, Logan , Utah, 
84322  	
\endaddress
\email torre\@cc.usu.edu\endemail
\thanks Research supported in part by NSF Grants DMS-\#9804833 and 
PHY-\#0070867 \endthanks
\date December 10, 2000\enddate
\endtopmatter

\document

\head 1. Introduction.\endhead

Inspired by Galois' theory of symmetry of polynomial equations, 
Sophus Lie (1842-1899) developed an analogous theory of symmetry for
differential equations. Besides this leading to the mathematics of
Lie groups and transformation groups, 
Lie's theory led to an algorithmic way
to find special explicit solutions to differential equations with
symmetry. These special solutions are called 
group invariant solutions and they constitute practically
every known explicit solution to the systems of non-linear partial 
differential
equations which arise in mathematical physics and 
differential geometry. For example 
the Schwarzschild solution to the vacuum Einstein 
equations, and the instanton solution to the
(self-dual) Yang-Mills equations are both group invariant solutions. 

Today the search for group invariant solutions is still a common
approach to explicitly solving non-linear partial differential 
equations.  
For instance, interesting 
examples of Einstein manifolds, harmonic maps
and Ricci-solitons with cohomogeniety one symmetry
groups can be found in \cite{4},\cite{7} and \cite{6}. 

An excellent introductory reference to Lie's theory
of group invariant solutions is \cite{8}. However in
this treatment (as in others \cite{3},\cite{9}) the theory is
developed to mainly study scalar partial differential
equations and does not apply to the standard equations
of mathematical physics. In fact one cannot even reproduce
the Schwarzschild solution from the techniques provided by these 
references.
The technical assumption made in these references which precludes 
these 
famous examples is known as {\it transversality}.  In 
this article, which is more or less a summary of \cite{1},
we provide a general method to find group invariant solutions for 
non-transverse group actions. The method also resolves 
the question of ``how many'' differential equations 
determine the group invariant solutions, a question which
does not arise for scalar equations.

\head 2. Notation and Preliminaries. \endhead

Let $\pi:E\to M$ be a fibre bundle with $n$-dimensional base, and 
$m$-dimensional fibre.  The map $\pi$ in local coordinates has the 
form 
$$
\pi(x^i, u^\alpha) = (x^i) \qquad i =1...n, \ \alpha = 1...m\ .
$$
The local coordinates $(x^i)$ on the base $M$, play the role of
the independent variables, while the local coordinates $(u^\alpha)$ on
the fibres will be the dependent variables. The
sections of $E$ will be denoted by
$$
S(E) = \{ s : M \to E \ , \quad \pi \circ s = I_M \}
$$
and $s\in S(E)$ takes the form $ u^\alpha = u^\alpha(x) $ in local 
coordinates.

A group $G$ acting on $E$ is said to act {\deffont projectably} (or 
by fibre preserving transformations) if there exists an action of $G$ 
on $M$ such that for all $g \in G$ the diagram
$$ 
\matrix   E & \mapright{g} & E \cr \mapdown{\pi} & & \mapdown{\pi} 
\cr M
 & \mapright{g} & M  \cr \endmatrix  
\tag{1}
$$
commutes. The notation $gp$ and $gx$ denotes the action of $G$ on $E$ 
and $M$ respectively.

Given a group $G$ acting projectably on $E$ ( and hence acting on $M$ 
) and
a point $x \in M$ then the {\deffont isotropy or stabilizer subgroup 
of} $x$ is
$$
G_x = \{ g \in G \ , | \ g x = x \} \ .
$$ 
For each $x\in M$ the subgroup  $G_x$ acts on the fibre $E_x$. That is
given $x \in M$ and $ g \in G_x $, and $ p \in E_x =\pi^{-1}(x)$ then 
as
a consequence of the commutative diagram (1) we find
$$
\pi( g p ) = g \pi( p ) = g x = x = \pi(p) 
$$
and hence $gp \in E_x$.

\head 3. Examples Part I. \endhead

\example{Example 1.a} Let $G = SO(3)$ act on $M = N = \reals^3 -0 $ 
in the usual way, and  let $E$ be the trivial bundle $\pi: M \times N 
\to M$. The action of $G$ on $E$
is
$$
R ({\bx}, {\bu}) = (R {\bx}, R{\bu}) \qquad {\bx} \in M\, , \ {\bu} 
\in N\, , \ R \in SO(3)\, .
$$
This action is projectable with the projected action of $G$ on $M$ 
being the one given.
\endexample

\example{Example 1.b} Let $G$ be any Lie group acting on manifolds 
$M$ and $N$.
Let $E$ be the trivial bundle $\pi: M \times N \to M$. Take the 
action of $G$ on $E$ to be the product action
$$
g ({\bx}, {\bu}) = (g {\bx}, g{\bu}) \qquad {\bx} \in M\, , \ {\bu} 
\in N\, , \ g \in G 
$$
which is projectable.
\endexample

\example{Example 2.a} Let $M$ be any differentiable manifold and let 
$G = diff(M)$ be the group of diffeomorphisms of $M$, and let $E$ be 
any tensor-bundle over $M$.  The action of $diff(M)$ on $M$ lifts to 
a natural action of $diff(M)$ on $E$, which in the case $E =TM$ and 
$\phi \in diff(M)$ is $\phi_*:TM \to TM$. These lifted actions are 
projectable.
\endexample

\example{Example 2.b} Let $\ G \subset diff(M)$ be any Lie group, 
then 
 the lift of this action to any tensor-bundle over $M$ as in example 
2.a is projectable.
\endexample

\head 4. Invariant Sections. \endhead

Given a group $G$ acting projectably on $E$ there is an induced
action of $G$ on $S(E)$ the sections of $E$. Given 
$ s \in S(E)$ and $ g \in G $ then the section $ g \cdot s $ is 
defined by 
$$
(g \cdot s ) (x) = g \, s ( g^{-1} x)  \  ,\qquad \forall \  x \in M .
$$

\definition{Definition 1} Let $G$ act projectably on $E$. A section
$ s \in S(E)$ is {\deffont $G$-invariant} if $ g \cdot s = s $ or
$$
g s(x) = s ( g x ) \ .
\tag{2} 
$$ 
\enddefinition
Denote by $S(E)^G$ the set of $G$-invariant sections of $E$, where of 
course $S(E)^G$ is the set of fixed points of the action of $G$ on 
$S(E)$.

\example{Example 1.b} (continued) The group $G$ acts on each
term separately in the product bundle $E = M \times N \to M$. In this
case the space $S(E)^G$ admits an alternate description. Given $ \phi 
\in C^\infty(M,N) $, the smooth maps from $M$ to $N$, we define a 
section $s_{\phi} : M \to E$ 
by $ s_{\phi}(x) = (x, \phi(x))$, which identifies
$$
S(E) = C^\infty(M,N) \ .
$$ 
Using this identification, a section $ s \in S(E)$ is invariant if and
only if
$$
g s_{\phi}(x) = ( g x , g \phi(x) ) = (g x, \phi(g x) ) = s_{\phi}(gx)
$$
or $ \phi(gx) = g \phi(x)$.  Thus the invariant sections
of $E$ are just the $G$-equivariant maps $ \phi : M \to N $
and for this reason group invariant solutions are sometimes
called equivariant \cite{7}.
\endexample

A fundamental problem in the theory of group invariant solutions
is how to parameterize the space $ S(E)^G$. Adding the following 
hypothesis, which is not often valid but whose definition will be
used later, makes this fairly easy.

\definition{Definition 2} A projectable group action $G$ on $E$,
is said to be {\deffont transverse} if for all $ p \in E$
$$
\pi( gp ) = \pi(p) \qquad {\text{\rm implies}}\quad  gp = p.
\tag{3}
$$ 
\enddefinition
Equivalently,  $G$ acts transversely on  $E$ if  the orbits of $G$  
in $E$  project diffeomorphically under  $\pi$ to the orbits  of $G$ 
in $M$. In local 
coordinates if the infinitesimal generators of the action of $G$ on 
$E$ are
given by
$$
\Gamma = \{ \xi^i_a ({\bx}) \partial_{x^i} + \phi^\alpha_a 
({\bx},{\bu}) \partial_{u^\alpha} \} \qquad  a = 1\dots dim \, G
$$
and if the action is transverse then
$$
rank \ [ \xi^i_a ({\bx}) ] = rank \ [ \xi^i_a ({\bx}), \phi^\alpha_a 
({\bx},{\bu}) ] \ .
$$
The invariant sections $S(E)^G$ for transverse actions are easily 
parameterized.
\proclaim{Theorem 1} Let $G$ be a Lie group acting projectably and
transversally on the bundle $\pi: E \to M$, and 
regularly on $M$ (so that M/G is a manifold).
Then $S(E)^G$ is in one-to-one correspondence with the sections of the
bundle $ E/G \to M/G$
\endproclaim
\demo{Proof} We will only show how a section $\tilde s : M/G \to E/G 
$ defines 
a section $s :M \to E$, and leave the rest as an exercise. Given 
$\tilde x \in M/G $ let $x \in M$ with $ \qM(x) = \tilde x$ where 
$\qM:M \to M/G$. 
First we claim there exists
a unique $p \in E$ such that 
$$ 
i] \ \qE(p) = \tilde s(\tilde x) \qquad ii] \  \pi(p) = x 
$$  
where $\qE:E \to E/G$. Suppose there exist $p,p'$ satisfying 
conditions $i],ii]$. By condition $i]$ $p' = g p$ for some $g \in G$,
which when used in condition $ii]$ along with transversality (3) 
implies $ p' = p $.  Therefore define $s(x) = p$, to get the 
appropriate section. See \cite{2} for more details. \qed
\enddemo

The following simple corollary demonstrates that transverse group 
actions
are unusual, and hence most examples of group actions on bundles
 are not transverse.
the following.
\proclaim{Corollary 1} Let $G$ act projectably on $E$, then
$G$ acts transversally if and only if for all $x \in M$,
$ G_x$ acts trivially on $E_x$.
\endproclaim
\medskip
For the examples we are interested in we need to find a way to 
parameterize the space $S(E)^G$ when the action of $G$ on $E$ may not 
be transverse. In order to do so, we use the following observation.  
Let $ s \in S(E)^G$ be 
an invariant section, let $x \in M$  and let $ g \in G_x$. The
invariant section condition (2) at $x$ gives
$$
g s( x) = s (gx) = s(x)
\tag{4}
$$
with the last identity coming from $g \in G_x$.  Therefore
if $s \in S(E)^G$ then $s(x) \in (E_x)^{G_x}$, the 
fixed point set of $G_x$ acting on $E_x$, which motivates the
following definition.

\definition{Definition 3} Let $G$ act projectably on $E$ and 
let $ \kappa_x(E) =  (E_x)^{G_x} $ and
$$
\kappa(E) = \bigcup_{x\in M} \kappa_x(E) \ .
$$
We call $ \pi : \kappa(E) \to M$ {\deffont the kinematic bundle}.
\footnote {
The set $\kappa(E)$ is not necessarily a manifold, but for our 
discussion we will assume it is and that $\pi : \kappa(E) \to M$ is a 
submersion.}
\enddefinition
A simple consequence of this definition and equation (4) is 
\proclaim{Corollary 2} Every $s\in S(E)^G$ factors through $\pi : 
\kappa(E) \to M$.
\endproclaim
\noindent
Note that this corollary does not say that every section of 
$\kappa(E)\to M$ is an element of $S(E)^G$. However the following is 
true,
\proclaim{Theorem 2} The subset $\kappa(E)\subset E$ is 
$G$-invariant, $G$ acts
transversally on $\pi :\kappa(E) \to M$ and $S(E)^G = S(\kappa(E))^G$.
\endproclaim
\demo{Proof} Let $p \in \kappa(E)$ with $x=\pi(p)$ and let $g\in G$, 
$h \in G_{gx}$. 
The subgroup $G_{gx} = g G_{x} g^{-1}$ therefore $h = g h' g^{-1}$ 
for some
$ h ' \in G_x$ and 
$$
h g p = g h' g^{-1} g p = g h' p = g p
$$  
so $ gp \in \kappa_{gx}(E)$. Transversality follows directly from
the definition of $\kappa(E)$. \qed
\enddemo

Theorem 2 along with Theorem 1 imply the next theorem.

\proclaim{Theorem 3} Suppose $\kappa(E)\to M$ is a bundle, and that 
$M/G$ is a manifold, then the space $S(E)^G$ of smooth invariant 
sections is in one-to-one correspondence with smooth sections of 
$\kappa (E)/G  \to M/G$.
\endproclaim

The situation is neatly summarized by the diagram,
\setbox6 =
\vbox{\hsize 18pc \strut
$\!\!\!\!\!\!\!\!\!$
\noindent
\underbar{Kinematic Reduction Diagram}

$
\qquad \qquad \matrix \cr {\tilde \kappa}(E) & \mapleft{\qk} & 
\kappa(E) & \mapright{\iota} & E \cr \mapdown{\pi} & & \mapdown{\pi} 
& & \mapdown{\pi} \cr {\tilM} & \mapleft{\qM} &  M & \mapright{id} & 
M  \cr \endmatrix
$
\strut}
$$
\boxit{\box6}
\tag{5}
$$
where $ {\tilde \kappa}(E) = \kappa(E)/G$ and $\tilM = M/G $, and  
every section of $\tilde \kappa (E)  \to \tilM$ lifts to an invariant
 section of $E\to M$.

It is worth emphasizing that the invariant sections of a bundle with 
a non-transverse group action are parameterized by the bundle $\tilde 
\kappa(E) \to \tilM$ whose base and {\it fibre} are typically of 
smaller dimension than $E \to M$.

\head 5. Examples Part II - Kinematic Reduction. \endhead

\example{Example 1.a} (continued) $G = SO(3)$, is acting on $M = N = 
\reals^3 -0 $ in the standard way, and $E$ is the trivial bundle 
$\pi: M \times N \to M$. 
The isotropy $G_{\bx}$ at $ \bx \in M$ is $SO(2)_{\bx}$, the 
rotations about the line containing $\bx$ and the origin. The set 
$\kappa_{\bx}(E) \subset E$ is
$$
\kappa_{\bx}(E) = \{ (\bx , \bu) \ | \ R \bu = \bu \ , \ R \in 
SO(2)_{\bx} \ \} \ 
$$
and so $ \bu \propto \bx $, or $ \bu = v \bx , \ v \in \reals $. The 
bundle
$\kappa(E)$ is a trivial line bundle with coordinates $(\bx , v)$ and 
the inclusion $\kappa(E) \to E$ given by $ (\bx , v ) \to (\bx, \bu = 
v \bx )$. The function $ r = \sqrt{x^2+y^2+z^2}$ defines a global 
coordinate on $M/G$. A coordinate description of the kinematic 
reduction diagram is
$$
\quad \qquad \matrix \cr (r,v) & \mapleft{\qk} & (\bx, v) & 
\mapright{\iota} & (\bx, \bu ) \cr \mapdown{\pi} & & \mapdown{\pi} & 
& \mapdown{\pi} \cr (r) & \mapleft{\qM} &  (\bx) & \mapright{id} & 
(\bx)  \cr \endmatrix 
\tag{6}
$$
where $ \iota(\bx,v) = (\bx, \bu = v \bx ) $.
Any invariant section $ v = v(r)$ of $ \tilde \kappa(E)\to \tilM$ 
(the left hand side of (6)) lifts to the invariant section $ u(\bx) = 
v(r) \bx $ of $E$. 
\endexample

\example{Example 2.b} (continued) Let $G = SO(3) \times ( Z_2 \sd 
\reals ) $ act on $(\reals^3 -0)\times \reals $ with local 
coordinates $(x,y,z,t)$ by
$$
(R,\tau,\epsilon) ( \bx ,t) = ( R \bx , \epsilon t + \tau ) \, . 
$$ 
The isotropy group at $(\bx,t_0)$ is $G_{(\bx,t_0)} = SO(2)_{\bx} 
\times Z_2 $ where the $Z_2$ action is
$$
(\bx , t) \to \left( \bx , \epsilon( t - t_0 )+t_0 \right) \ .
$$
Let $ E = T^*M \odot T^*M $, the bundle of symmetric tensors over $M$.
In this example the set $\kappa_{(\bx, t_0)}(E) $ which we wish to 
compute, is just the set of invariants of the symmetric tensor 
product of the dual of the linear isotropy representation.

In order to compute $ \kappa_{(\bx,t_0)}(E)$ we consider each term in 
the product group $G_{(\bx,t_0)}$ separately. For $ R\in 
SO(2)_{\bx}$, the condition
for $\gamma \in \kappa_{(\bx,t_0)}(E)$ is $ R^T \gamma R = \gamma $. 
Since $SO(2)_{\bx}$ is connected this condition can be written 
infinitesimally as
$$
X^T\gamma +  \gamma X =0  \quad  {\text {\rm where }} \ \ X = \left( 
\matrix  0 & z & - y &0 \cr -z & 0 & x &0\cr y & -x & 0&0 \cr 0 & 0 & 
0 & 0 \endmatrix \right) \ .
$$
Solving these equations gives 
$$
\gamma = A (x dx + y dy + z dz)^2 + B( dx^2+dy^2+dz^2) + C dt ^2 + D 
(x dx + y dy + z dz) \odot dt  
$$
where $ A,B,C,D \in \reals $. The constraint for the $Z_2$ part of
the isotropy is $ S^T \gamma S = \gamma $ where $ S = 
diag(0,0,0,\epsilon)$,
which implies $D = 0$. Therefore $ \kappa_{(\bx,t_0)}(E) = (A,B,C)$ 
and $\kappa(E)$ is
a rank $3$ vector subbundle of $E$. The kinematic reduction diagram is
$$
\matrix \cr (r,A,B,C) \!\!\! & \mapleft{\qk} & \!\! (\bx,t, A,B,C) & 
\mapright{\iota} & \!\!\! (\bx,t, \gamma ) \cr \mapdown{\pi} & & 
\mapdown{\pi} & & \mapdown{\pi} \cr (r) & \mapleft{\qM} &  (\bx,t) & 
\mapright{id} & (\bx,t) \cr \endmatrix 
\tag{7}
$$
where $r=\sqrt{x^2+y^2+z^2}$, and $\iota(\bx,t,A,B,C) =(\bx,t,\gamma 
= A(rdr)^2\!+\!B(dx^2+dy^2+dz^2) + C dt^2 )$.
Finally the sections $ (A(r),B=B(r), C=C(r))$ of $ \tilde \kappa (E)$ 
(the left side of diagram (7)) give rise to the $G$-invariant 
symmetric covariant $2$-tensors
$$
\gamma = A(r) (rdr)^2+B(r)(dx^2+dy^2+dz^2)+C(r) dt^2  \ .
$$
\endexample

\head 6. The Reduced Equations. \endhead

Let $ J^k(E)$ be the bundle of $k$-jets of sections of $E$ and  
let $G$ act projectably on $E$. Then $G$ acts on $J^k(E)$ 
by
$$
g \cdot \sigma = j^k( g\cdot s)(g x)
$$
where $\ \sigma = j^k(s)(x) \in J^k(E) $ and $j^k(s)$ is the 
$k^{th}$-jet of a section $s \in S(E)$.  A $k^{th}$-order 
differential operator $ \Delta $ is a section of a vector-bundle $ 
{\Cal D} \to J^k(E)$. A solution to the differential equations 
$\Delta = 0$ is a section $s : M \to E $ such that $\Delta \circ 
j^k(s) = 0$. 
Denote the solutions to $\Delta$ by the subset $S_\Delta(E) \subset 
S(E)$.

Let $G$ be a projectable group action on the bundle ${\Cal D} \to 
J^k(E)$, then $G$ is a symmetry group of $\Delta $ if $\Delta: J^k(E) 
\to {\Cal D}$ is a $G$-invariant section. It is easy to verify

\proclaim{Lemma 1} If $s \in S_\Delta(E)$ and $ g \in G$ a symmetry 
group of $\Delta$, then $g \cdot s \in S_\Delta(E)$. 
\endproclaim

In other words if $G$ is a symmetry group of $\Delta $ then the 
subset $S_\Delta(E)\subset S(E)$ is $G$-invariant. The fixed point 
set $S_\Delta(E)^G$ of the action of $G$ on $S_\Delta(E)$ leads to 
the definition

\definition{Definition 4} Let $G$ be a symmetry group for $\Delta$. 
A solution
$$
s\in  S_\Delta(E) \bigcap S(E)^G = S_\Delta(E)^G 
$$ 
is called a {\deffont $G$-invariant solution}.  
\enddefinition

We now determine a reduced differential operator $\widetilde \Delta $
whose solutions determine all the $G$-invariant solutions to 
$\Delta$. 
Every $G$-invariant solution to a differential equation
is an element of $S(E)^G$ which by Theorem 3 can be identified with
a section of $\tilde \kappa(E) \to \tilM$. Therefore 
the reduced differential operator will be a differential operator
on $J^k(\tilde \kappa(E))$ and define a section of some vector-bundle
$\widetilde {\Cal D} \to J^k( \tilde \kappa (E))$. 

 It is {\bf not} possible to directly apply the theory developed in 
section 4
to find the reduced differential operator and bundle $\widetilde 
{\Cal D}$ for $\Delta$. 
We need to introduce the so-called invariant jet space $Inv^k(E)$. 
The set $Inv^k(E) \subset J^k(E)$ is defined by
$$
Inv^k(E) = \{ \, \sigma \in J^k(E) \, , {\text{  where}} \, \sigma = 
j^k(s)(x) \ {\text{ for some}} \  s \in S(E)^G \}\ .
$$
We list some facts about invariant jet-space spaces.
\smallskip

\proclaim{Lemma 2} Let $G$ act projectably on $E$ then

a] $Inv^k(E) \subset J^k(E)$ is  $G$-invariant.

b] $G$ acts transversally on $Inv^k(E) \to M $.

c] Every $j^k(s):M \to J^k(E)$, for $ s \in S(E)^G$ factors through 
$Inv^k(E) $.
\endproclaim

\proclaim{Lemma 3} If $\kappa(E)$ is a bundle and $M/G$ is a manifold 
then

a] $Inv^0(E) $ is diffeomorphic to $\kappa(E)$.

b]  $Inv^k(E) \to M$ is equivalent to the pull-back bundle $\q^* 
J^k(\tilde \kappa(E))$ where $\qM:M \to M/G$.

\endproclaim

Suppose that the condition in Lemma 3 hold, let $ \qi : Inv^k(E) \to 
J^k(E)$ be the inclusion map, and let ${\Cal D}_\qi \to Inv^k(E)$ be 
the restriction (or pullback) of the vector-bundle ${\Cal D} \to 
J^k(E)$ to $Inv^k(E)$. Then 
we have,

\proclaim{Lemma 4} The group $G$ acts 
on the sub-bundle $i:Inv^k(E) \to J^k(E)$ and $\Delta $ defines an 
invariant section of the restricted bundle ${\Cal D}_{\qi} \to 
Inv^k(E)$. 
\endproclaim

At this point we apply the theory of section 4 to the
bundle ${\Cal D}_\qi \to Inv^k(E)$ to obtain the left side in the 
diagram
\setbox7 =
\vbox{\hsize 20pc \strut
$\!\!\!\!\!\!\!\!\!$
\noindent
\underbar{Dynamic Reduction Diagram}

$
\matrix \cr \kappa({\Cal D}_{\qi})/G & \mapleft{\q} & \kappa({\Cal 
D}_{\qi}) & \mapright{\iota} &{\Cal D}_{\qi} & \mapright{\qi} & {\Cal 
D} \cr \downarrow & & \downarrow & & \downarrow & & \downarrow \cr 
J^k(\tilde\kappa(E)) & \mapleft{\q} &  Inv^k(E) & \mapright{id} & 
Inv^k(E) & \mapright{i} & E \cr \endmatrix
$
\strut}
$$\boxit{\box7} \tag{8}$$

Theorem 3 states that the invariant sections of ${\Cal D}_i\to 
Inv^k(E)$ are in one to one correspondence with sections of 
$\kappa({\Cal D}_i)/G \to J^k(\tilde \kappa( E))$. Given an invariant 
differential operator $\Delta$ we call the operator $\widetilde 
\Delta$ obtained through this correspondence, the {\deffont reduced 
operator}.

\head 7. Examples Part III - Dynamic Reduction. \endhead

\example{Example 1.a} (continued) The Euler-Equations. Let $G = 
SO(3)$ act on $M = N = (\reals^3 -0) \times \reals $ in the standard 
way on the $\reals^3 -0 $ term, and let $E$ be the trivial bundle 
$\pi: M \times N \to M$.  We use $(\bx,t)$ and $(\bu,p)$ as 
coordinates on $M$ and $N$ respectively. The standard coordinates on 
$J^1(E)$ are $ (\bx,t,\bu, p\,;\, {u^i\!,}_{j}, u^i_t , p_{\,j}, 
p_t), \ i,j =1,2,3$. 

Let $ {\Cal D} = \reals^4 \times J^1(E)$ be a trivial rank 4 vector 
bundle. We
use $ (\bDel^i , \Delta^4) , \ i = 1,2,3$ for fibre coordinates, and 
take
for the action of $G$ on the fibres $ R(\bDel^i,\Delta^4 ) =
 ( R^i_j \bDel^j ,\Delta^4 )$, $R \in SO(3)$.  The Euler-equations 
are given by the vanishing of the differential operator $ \Delta : 
J^1(E) \to {\Cal D}$ 
$$
\eqalign {
\Delta ^i & = u^i_t + {u^i\!,}_j u^j + \delta^{ij}({p,}_{j}) \quad i 
=1...3 \cr \Delta^4 & =
{u^j\!,}_{j}  } 
$$
where $\delta^{ij}$ are the components of the three dimensional 
(contravariant) Euclidean metric in standard coordinates.

To find the reduced operator we first determine $ Inv^1(E) \to 
J^1(E)$ using a slight extension of diagram (6). The invariant 
sections of $E$ are simply seen to be $ S(E)^G= \{ \bu = v(r,t) \bx\ 
, \ p = p(r,t) \ \} $,
where $ r=\sqrt{x^2+y^2+z^2}$. The 
coordinates on $Inv^1(E)$ are then $(\bx,t,v,p,v_r,v_t,p_r,p_t)$ and
the inclusion $i:Inv^1(E) \to J^1(E)$ in coordinates is
$$
\eqalign{
& i:(\bx,t,v,p,v_r,v_t,p_r,p_t) \to \cr
& (\bx,t,\bu = v \bx ,p,{u^i,}_j=r^{-1}v_rx^ix^j + v \delta^i_j, 
u^i_t =v_t x^i, p_j=r^{-1}p_rx^j, p_t=p_t)\ .}
$$
\smallskip
\noindent
To find the reduced operator we restrict $\Delta$ to $Inv^1(E)$
which gives
$$
\eqalign {
\Delta^i_{Inv} & = \left( v_t +v (v + r v_r) +r^{-1} p_r \right) 
x^i\cr
\Delta^4_{inv} & = 3v+r v_ r \ .} 
$$
From this $\kappa({\Cal D}_i) $ is a rank 2 bundle, and the 
components of the reduced operator are
$$
\tilde \Delta =  \left( v_t +v (v + r v_r) +r^{-1} p_r,  3v+r v_ r 
\right) \ .
$$

\endexample

\example{Example 2.b} (continued) The Schwarzschild solution. Here $ 
G = SO(3) \times Z_2 \sd  \reals  $, 
   $ M = ( \reals ^3 -{0})\times \reals $, and $ E = T^*M \odot T^*M 
$.
Using $ \Pi : J^2(E) \to M$, let ${\Cal D} = \Pi^* E$ be the 
pull-back bundle and let $G$ act on ${\Cal D}$ in the natural way.  
In this example the differential operator $ \Delta :J^2(E) \to {\Cal 
D}$ is
$$
\Delta   = R_{ij} dx^i \odot  dx^j
$$
where $R_{ij}$ are functions of the second derivatives of a symmetric 
covariant two tensor,  and $\Delta = 0$ are the vacuum Einstein 
equations, or the Ricci-flat conditions.
\medskip
The differential operator $\Delta$ takes values in $T^*M \odot T^*M$, 
so the bundle $\kappa({\Cal D}_i)$ will have the same structure as 
$\kappa(E)$ (see (7)). Thus restricting $\Delta$ to $Inv^2(E)$ we find
$$
\Delta_{Inv}  = 
{\widetilde \Delta}_A(x dx \!+\! y dy\! +\! z dz)^2 \!+\! {\widetilde 
\Delta}_
B (dx^2\!+\!dy^2\!+\!dz^2)\!+\!
{\widetilde \Delta}_C dt^2 \ ,
$$
where ${\widetilde \Delta}_A, {\widetilde \Delta}_B$, and 
${\widetilde \Delta}_C$ are the components of the reduced operator. 
Solving the reduced equations $\widetilde \Delta_{A,B,C} =0$ leads to 
the Schwarzschild solution.

\endexample

\head 8.  Quotients with boundary. \endhead

In a number of applications, 
such as the cohomogeniety one reductions of harmonic map and Einstein 
equations \cite{4},\cite{7}, the quotient 
$M/G$ is a manifold with boundary and consequently Theorem 3 does not 
hold and a general theory to parameterize the space $S(E)^G$ is 
unknown. 

\example{ Example 4 } Let $  G = SO(2) $, $  M = N = S^2 \subset 
\reals^3 $, $  E = M\times N \to M $ where the action is by rotation 
about the $z$ axis on both $S^2$.  The action is not
transverse because at the poles $(0,0,\pm 1) \in M$ we have $ G_ 
{(0,0,\pm 1) } = SO(2) $ and this acts non-trivial on the $S^2$ 
fibre. The isotropy
condition at $(0,0,\pm 1)$ leads to 
$$
\kappa_{(0,0,\pm 1) }(E) = (0,0,\pm 1) \ .
\tag{9}
$$
Therefore any $SO(2)$ equivariant map from $S^2$ to
$S^2$ must take the poles to the poles. The quotient $M/G=[-1,1]$ and 
condition (9) would lead to constraints on the boundary conditions 
for any group invariant solution having this symmetry. See \cite{1} 
for an explicit application to harmonic maps.
\endexample

The above example shows that we can use the $\kappa$ ``functor''
to obtain information about the space of invariant sections
of $E$ when $M/G$ is a manifold with boundary. 
Using the projection maps $\pi^k : J^k(E) \to M$ we can construct 
$\kappa(J^k(E))$ where $ \kappa_x(J^k(E)) = \left( \, J_x^k(E) \, 
\right)^{G_x}$. It is clear that $ Inv^k(E) \subset \kappa(J^k(E)) $ 
and so
$\kappa(J^k(E))$ constrains the $k$-jet of any invariant section. 
It is also possible to define $\kappa(J^k(E))$ inductively using the
projection maps $\pi^{k}_{k-1}: J^{k}(E) \to J^{k-1}(E)$. Let
$\sigma \in \kappa(J^{k-1}(E))$ then
$$
\kappa_\sigma(J^{k}(E)) = \left( (\pi^{k}_{k-1})^{-1}(\sigma)\right) 
^{G_\sigma} \ . 
\tag{10}
$$
Note that $\kappa(J^0(E)) = \kappa(E)$.

\example{ Example 5 } Let $ G = SO(n) $, $  M = \reals^n $, and $  E 
= M \times \reals $ so $S(E)^G$ is the space of $SO(n)$-invariant 
functions on $\reals^n$, and
$ M/G = [0,\infty) $. Unlike the case when $M/G$ is manifold, the
space $S(E)^G$ cannot be identified with the smooth functions $f : 
M/G \to \reals$. 

Let $\sigma \in J^{k-1}(E)$ with $\pi^{k-1}(\sigma) = \bo $ the 
origin $\reals^n$.  We identify 
$$
(\pi^{k}_{k-1})^{-1} ( \sigma ) = \odot^{k} (\reals^n) 
$$
which represent the $k^{th}$ order coefficients of the Taylor 
polynomial
of a smooth-functions $f$ defined in a neighbourhood of $\bo$.
We compute $\kappa_{\bo}(J^k(E))$ using (10). The stabilizer at $\bo$ 
is $G_{\bo} = SO(n)$ and for $ \sigma \in \kappa(J^{k-1}(E)) $ with $ 
\pi^{k-1}(\sigma) = \bo $ we have
$$
\kappa_\sigma(J^k(E)) = \left( \odot^k \reals^n \right)^{SO(n)}, 
$$
where the $SO(n)$ action is the $k^{th}$ order symmetric tensor 
product
of the standard action of $SO(n)$ on $\reals^n$. It is well known 
\cite{11} that
$  \left( \odot^k \reals^n \right)^{SO(n)} = 0 $ for $k$ odd. These 
necessary conditions are easily derived in even dimensions from the 
fact that if $f \in S(E)^G$ then $f(-\bx) = f(\bx)$, or $f(r)$ is 
even. 

For examples where these conditions can be used to guarantee 
smoothness see \cite{5} or \cite{10}.
\endexample

\head References. \endhead

\refstyle{A}

\enddocument